\documentclass[sigconf]{acmart}

\usepackage{amsmath}
\usepackage{amsfonts}
\usepackage{algorithm}
\usepackage{algpseudocode}
\usepackage{array}
\usepackage{makecell}
\usepackage[table]{xcolor}

\copyrightyear{2026}
\acmYear{2026}
\setcopyright{cc}
\setcctype{by}
\acmConference[ICCAD '26]
  {IEEE/ACM International Conference on Computer-Aided Design}
  {November 08--12, 2026}
  {San Jose, CA, USA}
\acmBooktitle{IEEE/ACM International Conference on Computer-Aided Design
  (ICCAD '26), November 08--12, 2026, San Jose, CA, USA}
\acmDOI{10.1145/3831252.3834056}
\acmISBN{979-8-4007-2873-0/2026/11}

\begin{document}

\title{Vector Field Transforms: Structured Hyperdimensional Memory
for Spatio-Temporal Reasoning}

\author{William Chung}
\affiliation{%
  \institution{University of California, Irvine}
  \city{Irvine}
  \state{CA}
  \country{USA}
}
\email{chungwy1@uci.edu}

\author{Andrew Ding}
\affiliation{%
  \institution{University of California, Irvine}
  \city{Irvine}
  \state{CA}
  \country{USA}
}
\email{ading2@uci.edu}

\author{Hansen Jin Lillemark}
\affiliation{%
  \institution{University of California, San Diego}
  \city{San Diego}
  \state{CA}
  \country{USA}
}
\email{hlillemark@ucsd.edu}

\author{Mohsen Imani}
\affiliation{%
  \institution{University of California, Irvine}
  \city{Irvine}
  \state{CA}
  \country{USA}
}
\email{m.imani@uci.edu}

\renewcommand{\shortauthors}{Chung et al.}
\settopmatter{authorsperrow=4}

\begin{abstract}
Vector Symbolic Architectures (VSAs) are attractive for edge intelligence because their fixed-width distributed representations and simple algebraic operations map naturally to low-power vector hardware. However, conventional VSA memory superposes information, such as object, position, and time, into a single global vector causing interference to grow with scene complexity and trajectory length. We propose Vector Field Transforms (VFT), a symbolic spatio-temporal memory that replaces the flat global superposition with a spatially indexed field of local VSA memories. On a local clutter experiment, VFT achieves 100.0\% top-1 accuracy compared with 20.3\% for MAP-VSA and 37.3\% for FHRR-VSA. For a temporal recall benchmark, VFT achieves 74.8\% accuracy while flat temporal VSA reaches only 5.8\% and 10.6\%. Additionally, we map VFT's transport, write, and query kernels to a proposed programmable RVV processor. A DMA-assisted field-shift engine reduces VFT step latency by 1.40--1.45$\times$, including a reduction from 55.1~$\mu$s to 38.6~$\mu$s on the egomotion configuration. These results establish VFT as a practical symbolic memory architecture for spatio-temporal edge workloads and a strong match for programmable vector processors.
\end{abstract}

% CCS concepts
\begin{CCSXML}
<ccs2012>
   <concept>
       <concept_id>10010147.10010178.10010187.10010197</concept_id>
       <concept_desc>Computing methodologies~Spatial and physical reasoning</concept_desc>
       <concept_significance>500</concept_significance>
       </concept>
   <concept>
       <concept_id>10010583.10010786.10010787.10010788</concept_id>
       <concept_desc>Hardware~Emerging architectures</concept_desc>
       <concept_significance>500</concept_significance>
       </concept>
   <concept>
       <concept_id>10010520.10010521.10010528.10010534</concept_id>
       <concept_desc>Computer systems organization~Single instruction, multiple data</concept_desc>
       <concept_significance>300</concept_significance>
       </concept>
 </ccs2012>
\end{CCSXML}

\ccsdesc[500]{Computing methodologies~Spatial and physical reasoning}
\ccsdesc[500]{Hardware~Emerging architectures}
\ccsdesc[300]{Computer systems organization~Single instruction, multiple data}

\keywords{Vector Symbolic Architectures, Hyperdimensional Computing, RISC-V, Vector Processors, Spatio-Temporal Memory, Edge AI}

\maketitle
\section{Introduction}
Edge systems such as mobile robots, autonomous vehicles, and AI sensors must do more than classification. They must maintain a compact, query-able memory while operating under tight power and memory budgets. In many such settings, the core challenge is spatio-temporal memory. An agent must remember what it has seen, where it was seen relative to the agent, and how that memory should change as the agent moves. Classical neural world models can address this problem with structured latent memories~\cite{collu2024slot, po2025long, lillemark2026flow}, but their compute and memory cost makes them difficult to deploy on highly constrained edge settings.

\begin{figure}
    \centering
    \includegraphics[width=\columnwidth]{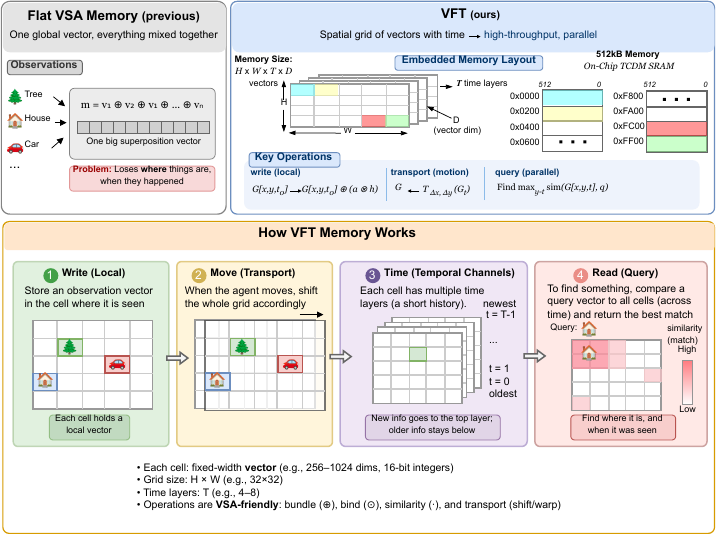}
    \caption{Vector Field Transforms (VFT) organizes memory into a spatially-indexed $H \times W \times K \times D$ local vector field . Core operations include: (1) Write, (2) Moves, (3) Time, and (4) Reads. VFT utilizes VSA-friendly bundling ($\oplus$), binding ($\otimes$), similarity ($s$), and shift-based transport.}
    \label{fig:overview}
     \vspace{-5mm}
\end{figure}

Vector Symbolic Architectures (VSAs), also known as Hyperdimensional Computing \cite{kanerva2009hyperdimensional},  offers a fundamentally different approach to edge intelligence. By representing symbols as high dimensional distributed vectors and composing them through simple algebraic operations, VSA systems achieve noise-tolerant, fixed-width computation that maps naturally to low-precision arithmetic and data-parallel execution. Crucially, memory and computation are unified through the same operations on distributed representations, with no pointer-based addressing or centralized data structures. This makes VSA attractive for power-constrained edge inference.

However, conventional hyperdimensional memories architectures operate through a single flat superposition where all bound items are accumulated into one global memory vector~\cite{kanerva1988sparse, plate2003HRR}. This is effective for associative lookup and prototype-based classification~\cite{ge2020classification}, but is poorly matched for embodied spatio-temporal tasks. When object, position, and time bindings are repeatedly superposed into one memory, interference grows with both scene complexity and trajectory length, and the representation provides no explicit mechanism to update stored content under viewpoint changes.

This paper proposes Vector Field Transforms (VFT), a new HDC computing paradigm that replaces flat global superposition with a spatially indexed field of local memory cells updated by an operator and temporal age channels. Rather than bundling all observations into a single global vector, it maintains a structured memory that evolves under agent motion while preserving the simple algebraic operations that make HDC attractive for edge hardware. Because VFT is dominated by field shifts, weighted transport, local writes, and parallel similarity queries rather than large matrix multiplications, it naturally maps onto HyperV~\cite{cicc2026}, a proposed out-of-order RVV vector processor for short and medium length vector workloads. This makes VFT a practical hardware-aligned VSA for spatio-temporal symbolic memory workloads.

We make the following contributions:
\begin{enumerate}
    \item We introduce VFT, a new hyperdimensional computing paradigm that replaces flat superposition with a spatially indexed, motion-transported field of local memories.
    \item We map VFT's core kernels, field shift, weighted transport, local write, and query, to the programmable RVV vector processor architecture. A DMA-assisted field-shift engine removes the shift kernel from the critical path and yields a consistent 1.40--1.45$\times$ step-speedup, improving the egomotion configuration from 55.1~$\mu$s to 38.6~$\mu$s per VFT step.
    \item Across collision-heavy spatial memory, noisy egocentric transport, and temporal recall benchmarks, VFT substantially outperforms flat VSA under both MAP and FHRR encodings, confirming the importance of structured memory organization in spatio-temporal tasks. VFT also retains 95\% of its 64-bit accuracy using only 4-bit storage.
\end{enumerate}

\section{Related Works}
\subsection{Vector Symbolic Architectures and Symbolic Memory}
Vector Symbolic Architectures (VSAs), including HRR~\cite{Plate1995HRR} and MAP \cite{gayler1998multiplicative}, represent symbols as high-dimensional distributed vectors and simple algebraic operations such as bundling and binding~\cite{kanerva2009hyperdimensional}. This makes VSA-based methods attractive for associative memory, reinforcement learning, and robotics~\cite{ni2024efficient, hd_5, neubert2019introduction}, yet the limitation of VSA memory is well known~\cite{yeung2025cognitive}. As more items are superposed into a single memory, retrieval quality degrades due to interference between symbols in the set. In settings where an agent continually accumulates object, location, and temporal bindings, this interference grows rapidly with both scene complexity and trajectory length.

Several extensions improve how VSAs encode spatial or temporal structure. Fourier-based schemes support continuous coordinates and permutation-based encodings have been used for sequential encoding \cite{frady2022computing, choo2010spiking, komer2019neural, hd_4, ge2020classification, kanerva2009hyperdimensional}. However, these methods primarily encode structure within a flat superposition rather than maintaining an explicit spatio-temporal memory. As a result, memory updates must be handled indirectly through rebinding or re-encoding, which becomes increasingly inefficient and interference-prone as the stored scene grows. VFT addresses this limitation by introducing an explicit structured memory transition, allowing symbolic memory to evolve under motion rather than relying solely on repeated bundling and binding.

\subsection{Learned Spatial Transforms and Structured Memory}
Neural architectures have similarly introduced structured spatial state through differentiable warping~\cite{jaderberg2015spatial}, spatial recurrence~\cite{shi2015convolutional}, and explicit map memories~\cite{parisotto2017neural}. Rather than collapsing spatially distinct observations into a single global representation, these approaches preserve the local organization of features so that memory can be updated selectively. Spatial Transformer Networks geometrically align feature maps, while ConvLSTM and Neural Map maintain recurrent or addressable spatial states across observations. FloWM further transports a persistent latent memory by modeling self-motion and external object motion as structured flows~\cite{lillemark2026flow}. Additionally, Hierachical Equivariant Sparse Coding~\cite{shewmake2023visual} has tried learning composable structures and seen success on toy benchmarks. Together, these methods demonstrate that explicitly organizing and transporting spatial state can preserve information across viewpoint changes and long rollouts. However, their dense neural representations require learned updates and relatively expensive tensor operations, limiting their suitability for highly constrained edge systems. However, recently learnable VSA models~\cite{chung2026geometric} have shown some promise in modeling spatial transforms on grid-world environments, but haven't seen scalable success. VFT adopts the same principle of an explicitly evolving spatial state while implementing its memory through local hypervectors and element-wise VSA operations.

\subsection{Hyperdimensional Computing Hardware}
The hardware literature on VSA and HDC has largely developed between accelerators for associative inference, and domain-specific extensions for structured workloads. VSA is most often known as HDC when applied in the hardware domain. Early work focused on making HDC classification practical in hardware by accelerating encoding, bundling, and similarity search over class prototypes. For example, FACH \cite{imani2019fach} reduces the cost of reasoning by restructuring the similarity computation to avoid most multiplications for efficient implementations on FPGAs. Neuromorphic VSA designs exploit the low-precision, element-wise nature of symbolic operations~\cite{abhijith2021neuromorphic}. HyperSpike extends this direction to event-driven sensing by coupling an SNN front-end with an VSA classifier and dedicated hardware support, demonstrating that VSA can serve as a robust and efficient decision layer for spiking pipelines \cite{morris2022hyperspike}. 

RelHD~\cite{kang2022relhd} and RelHDx~\cite{kang2025relhdx} map node features and graph structure into hyperdimensional representations suitable for efficient hardware realization, while HyperGRAF pushes this further with an FPGA-oriented accelerator specialized for HDC-based graph reasoning \cite{chen2023hypergraf}. More recently, \cite{chung2025robust} proposes learning the noise distribution to deal with any non-linearities resulting in emerging hardware designs, making HDC more robust and applicable on different devices. Other recent ASIC-style designs further confirm that symbolic representations can deliver attractive robustness efficiency tradeoffs in edge settings \cite{datta2023hdbinarycore, chung2026vector}. However, most prior designs are organized around fixed-function pipelines for prototype matching, graph aggregation, or application-specific inference. No prior works have targeted accelerating VSA memory systems designed for spatio-temporal tasks which require indexing, querying, and updating over time, which motivates a more programmable vector design.

\subsection{Programmable RISC-V Vector Processors for Edge AI}
A complementary line of work has explored programmable vector processors for data-parallel edge workloads. Recent RISC-V vector designs such as Hwacha \cite{hw}, Ara2 \cite{ar}, Spatz \cite{tc25}, and Maestro \cite{ma} show that vector ISAs can offer a useful balance of efficiency and flexibility across embedded applications. General-purpose vector processors provide a compelling and practical alternative for VSAs, as their scalable vector execution, support for element-wise arithmetic, and efficient reduction operations naturally align with VSA workloads. The use of vector processors as a flexible and silicon-proven substrate for VSA acceleration avoids the rigidity of fixed-function accelerators while retaining high efficiency. HyperV~\cite{cicc2026} targets short and medium vector execution through an out-of-order RVV microarchitecture with beat-multiplexed execution and a monolithic vector register file. This design aligns closely with VFT, whose core operators consist of row-wise field shifts, weighted accumulation over a small transport basis, local symbolic writes, and reduction-based query scoring. Recently, \cite{ding2026algorithm} has shown how HDC models can be mapped onto HyperV and shows high energy efficiency and good latency compared to CPU and GPU baselines. Table~\ref{tab:comparison} compares HyperV against a range of state of the art RISC-V vector processors.

\begin{table*}[t]
\caption{Architectural comparison of programmable RISC-V vector
processors. HyperV measurements are from fabricated silicon at
0.9\,V and 190\,MHz; the proposed VFT Field Shift DMA Engine is
evaluated separately in Tables~\ref{tab:kernel_perf}
and~\ref{tab:dma_perf}.}
\label{tab:comparison}
\centering
\small
\renewcommand{\arraystretch}{1.08}
\setlength{\tabcolsep}{4pt}

\begin{tabular*}{\textwidth}{
@{\extracolsep{\fill}}
l c c c c c c
}
\toprule
\textbf{Metric}
& \textbf{Hwacha~\cite{hw}}
& \textbf{Ara2~\cite{ar}}
& \textbf{Spatz~\cite{tc25}}
& \textbf{Maestro~\cite{ma}}
& \textbf{HyperV~\cite{cicc2026}} \\
\midrule

Technology
& 16\,nm
& 22\,nm
& 12\,nm
& 65\,nm
& 28\,nm \\

ISA
& RV64XHwacha
& RV64GCV1.0
& RV64GCV1.0
& RV32IMCV1.0
& RV32IMAFCV1.0 \\

VRF organization
& Lane
& Lane
& Monolithic
& Monolithic
& Monolithic \\

Parallel organization
& 9 lanes
& 4--16 lanes
& 8 lanes
& 4 VPUs + VTU
& 4 VPUs \\

Vector scheduling
& In-order
& In-order
& In-order
& In-order
& Out-of-order \\

Die area (mm$^2$)
& 4.41\newline(VPU only)
& 4.47
& 0.74
& 12.0
& 2.7 \\

Frequency
& 339\,MHz--1.44\,GHz
& 860\,MHz--1.35\,GHz
& 1.26\,GHz
& 90--210\,MHz
& 188--200\,MHz \\

\bottomrule
\end{tabular*}
\end{table*}

\section{Vector Symbolic Architectures}
\subsection{Background and Encodings}
VSAs represent symbols as high-dimensional hypervectors and compose them using simple algebraic operations such as bundling and binding. In large dimension $D$, randomly sampled hypervectors are approximately orthogonal, enabling compact associative memories and structured representations. Let $\mathbf{a},\mathbf{b}\in\mathbb{V}^D$. Bundling aggregates items by element-wise addition,
\begin{equation}
\mathbf{m}=\sum_{j=1}^{n}\mathbf{v}_j,
\end{equation}
while binding associates symbols into a new hypervector. A standard associative memory is
\begin{equation}
\mathbf{M}=\sum_{i=1}^{n}\mathbf{e}_i\odot\mathbf{p}_i,
\end{equation}
where $\mathbf{e}_i$ denotes object identity and $\mathbf{p}_i$ denotes an attribute such as position, role, or time.

We consider two common instantiations. In MAP, $\mathbf{v}\in\{-1,+1\}^D$, binding is element-wise multiplication, and similarity is the real inner product
\begin{equation}
s_{\text{MAP}}(\mathbf{u},\mathbf{v})=\mathbf{u}^{\top}\mathbf{v}.
\end{equation}
In FHRR, $\mathbf{v}\in\mathbb{C}^D$ is a complex unit-phasor vector, binding is element-wise complex multiplication, and similarity is
\begin{equation}
s_{\text{FHRR}}(\mathbf{u},\mathbf{v})=\Re\{\mathbf{u}^{H}\mathbf{v}\}.
\end{equation}
MAP is attractive for hardware due to simple sign multiplication and integer accumulation, while FHRR provides added benefits in continuous or phase-based representations.

\subsection{Flat VSA Memory and Its Limitations}
Using either encoding, object-location pairs can be stored with a position codebook $\mathbf{p}_{x,y}$ as
\begin{equation}
\mathbf{M}=\sum_{i=1}^{N}\mathbf{e}_i\odot\mathbf{p}_{x_i,y_i}.
\end{equation}
Retrieval is performed by scoring candidate positions and selecting the best match:
\begin{equation}
(\hat{x}_i,\hat{y}_i)=\arg\max_{x,y}\; s\!\left(\mathbf{M},\mathbf{e}_i\odot\mathbf{p}_{x,y}\right).
\end{equation}
Temporal information can be added similarly by binding a time code,
\begin{equation}
\mathbf{M}=\sum_{i,\tau}\mathbf{e}_i\odot\mathbf{p}_{x_i^\tau,y_i^\tau}\odot\mathbf{t}_{\tau}.
\end{equation}

This formulation is compact and compositional, but it remains \emph{flat}: all bindings are superposed into a single global vector. As memory grows, interference increases, and in embodied settings there is no explicit mechanism to transport stored content under viewpoint changes.

\section{Vector Field Transforms}
\label{section:VFT}
VFT preserves VSA bundling and binding, but replaces the single global memory with a spatial field of local memory cells. This allows memory to evolve under motion without global rebinding.

\subsection{State and Local Write Rule}
The VFT state at time $t$ is
\begin{equation}
M_t \in \mathbb{V}^{H\times W\times K\times D},
\end{equation}
where $\mathbb{V}=\mathbb{R}$ for MAP and $\mathbb{V}=\mathbb{C}$ for FHRR, $(H,W)$ defines the spatial field, $K$ is the number of temporal age channels, and $D$ is the hypervector dimension. Each cell $M_t[x,y,k,:]$ is a local VSA memory vector.

Given an observation of object $i$ at egocentric cell $(x,y)$, VFT writes to the current-age channel by
\begin{equation}
M_t[x,y,0,:]\leftarrow M_t[x,y,0,:]+e_i.
\label{eq:vft_write_short}
\end{equation}
Thus, location is represented by cell index rather than by binding to a global position code, and collisions remain local to each cell.

\subsection{Transport and Temporal Aging}
When the agent moves, the egocentric field must be shifted in the opposite direction to remain aligned with the current viewpoint. Let $\mathcal{B}=\{\text{stay},\text{up},\text{down},\text{left},\text{right}\}$ denote the set of basis shifts. Given observed motion $\hat{u}_t$, transport is
\begin{equation}
\widetilde{\mathbf{M}}_t
=
\sum_{b\in\mathcal{B}}
\alpha_{\hat{u}_t}[b]\,
\mathrm{Shift}_b(\mathbf{M}_{t-1}),
\label{eq:vft_transport}
\end{equation}
where $\alpha_{\hat{u}_t}$ is either a deterministic inverse motion or a distribution over plausible shifts under uncertainty.

To support temporal recall, each cell stores $K$ age channels. After transport, the field is aged before new writes:
\begin{align}
\mathbf{M}'_t[:,:,0,:] &= \widetilde{\mathbf{M}}_t[:,:,0,:], \\
\mathbf{M}'_t[:,:,k,:] &=
\rho\,\widetilde{\mathbf{M}}_t[:,:,k,:]
+
(1-\rho)\,\widetilde{\mathbf{M}}_t[:,:,k-1,:],
\label{eq:vft_age_short}
\end{align}
for $k=1,\dots,K-1$, followed by the local update in \eqref{eq:vft_write_short}. This yields a compact multiscale trace of recent history without having to bind a separate time vector into every observation.

\subsection{Querying}
Queries are performed by local cleanup over the field. For object $i$ and age channel $k$, the similarity heatmap is
\begin{equation}
S_{i,k}(x,y)=s\!\left(\mathbf{M}_t[x,y,k,:],\mathbf{e}_i\right),
\end{equation}
and the predicted location is
\begin{equation}
(\hat{x}_i,\hat{y}_i)=\arg\max_{x,y} S_{i,k}(x,y).
\end{equation}
More generally, a relative age $\tau$ can be queried by aggregating channels:
\begin{equation}
\bar{\mathbf{M}}_t^{(\tau)}[x,y,:]
=
\sum_{k=0}^{K-1}
w_\tau[k]\,
\mathbf{M}_t[x,y,k,:],
\end{equation}
followed by
\begin{equation}
S_{i,\tau}(x,y)=s\!\left(\bar{\mathbf{M}}_t^{(\tau)}[x,y,:],\mathbf{e}_i\right).
\end{equation}

Compared with flat VSA memory, VFT converts retrieval from global cleanup over one superposed vector into local cleanup over a transported spatial field, reducing interference and making memory updates compatible with motion and temporal aging.

\section{Hardware: HyperV}

\begin{figure}
    \centering    \includegraphics[width=1.0\columnwidth]{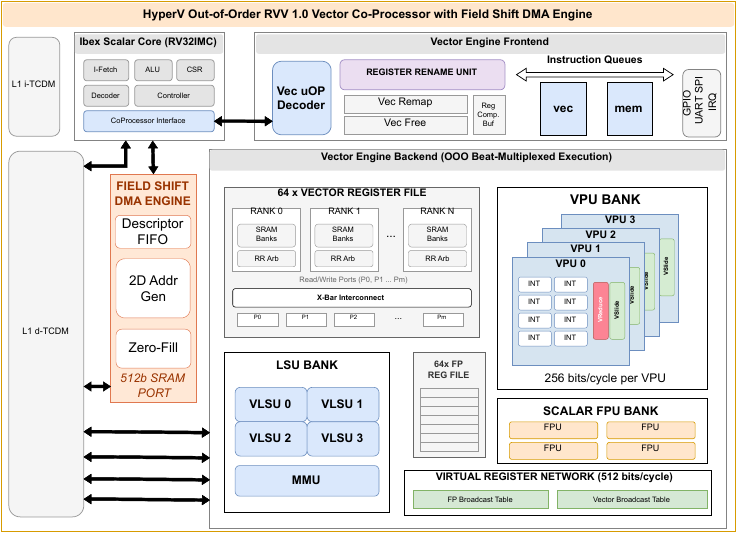}
    \caption{Architecture overview of the HyperV out-of-order RVV 1.0 vector co-processor and VFT hardware modifications. A dedicated Field Shift DMA engine offloads structured data movement to operate concurrently with arithmetic kernels.}
    \label{fig:arch}
     % \vspace{-3mm}
\end{figure}

\subsection{HyperV Micro-architecture Overview}
HyperV is an out-of-order RVV 1.0 vector co-processor fabricated in TSMC 28\,nm CMOS, optimized for edge AI inference, achieving 130 HP-FLOPS/W of compute efficiency. Tightly coupled to a lowRISC Ibex scalar core (RV32IMC), it offloads data-parallel operations while the scalar core handles control flow. The architecture has three innovations:
\begin{enumerate}
    \item \textbf{Beat-Multiplexed Execution:} Vector instructions are decomposed into micro-operations and dispatched out-of-order to multiple Vector Processing Units (VPUs), each processing 256 bits per cycle (one beat).
    \item \textbf{Monolithic Ranked VRF:} HyperV implements a 64-entry physical vector register file organized into ranks, avoiding the inter-lane routing overhead of traditional lane-based designs. A Register Rename Unit (RRU) eliminates WAW/WAR hazards, maximizing utilization for the medium vector lengths ($D=64$ to $256$) prevalent in VFT.
    \item \textbf{Tightly Coupled Memory System:} With 576\,KB of on-chip SRAM, multiple Vector Load-Store Units (VLSU) operate concurrently with compute units. The Virtual Register Network (VRN) provides 512\,bits/cycle of interconnect bandwidth, ensuring the VPU is rarely data-starved.
\end{enumerate}

This architecture serves as the basis of the proposed VFT accelerator. We implemented two modifications to the HyperV architecture to better support VFT workloads. The first modification incorporates a dedicated Field Shift DMA engine (Section~\ref{subsec:dma}) to accelerate the basis shift kernel. Secondly, the VPU's are parameterized to support int8, int16, and int32 data types, enabling efficient quantized VFT implementations.

\subsection{Kernel Mapping and RVV Co-Design}
VFT is an ideal workload for HyperV. Unlike DNNs dominated by monolithic GEMM, VFT relies on medium-scale, regular vector loops that balance memory bandwidth and compute. We decompose one VFT step (transport, write, query) into four RVV kernels:

\textbf{K1. Basis Shift Kernel.}
The basis shifts in Eq. \eqref{eq:vft_transport} copy (D)-element cell vectors to displaced field locations and zero the vacated boundaries. In software, this maps to unit-stride vector loads (\texttt{VLE16.V}) and stores (\texttt{VSE16.V}), but since no arithmetic is performed, it wastes VPU execution slots. We therefore offload this to the Field Shift DMA Engine described in Section~\ref{subsec:dma}.

\textbf{K2. Basis Mix Kernel.}
This kernel computes the transport $\widetilde{M} = \sum_{b} \alpha_{\hat{u}}[b] \cdot \text{Shift}_b(M)$. The weights $\alpha[b]$ are quantized to int16 ($\alpha_q = \lfloor 256\alpha + 0.5\rfloor$). For each cell, widening multiply-accumulate operations (\texttt{VWMACC.VX}) multiply the int16 vector by scalar $\alpha_q[b]$, accumulating into 32-bit registers ($\text{LMUL}{=}1 \rightarrow \text{LMUL}{=}2$). After all five bases are accumulated, a narrowing saturating clip (\texttt{VNCLIP.WX}) with right shift by 8 produces the int16 result. The Ibex core broadcasts coefficients via the scalar FPU bank, and out-of-order issue with vector chaining allows the VPU to begin accumulating while the VLSU loads the next element, hiding memory latency.

\textbf{K3. Local Write Kernel.}
Writing the code book vector $e_i$ into the spatial field (Eq.~\eqref{eq:vft_write_short}) is a single vector add: the cell vector is loaded (\texttt{VLE16.V}), the token added element-wise (\texttt{VADD.VV}), and the result stored (\texttt{VSE16.V}). Since MAP codebook entries are bipolar ($\pm 1$), no multiplication is needed.

\textbf{K4. Query Kernel.}
Evaluating similarity requires $H \times W$ independent dot products between each cell and the query token $e_i$, implemented as a widening multiply (\texttt{VWMUL.VV}, int16 $\times$ int16 $\rightarrow$ int32) followed by a reduction sum (\texttt{VREDSUM.VS}). The scalar result is extracted (\texttt{VMV.X.S}) and compared against a running maximum on the Ibex core. The independence of these reductions allows the OOO pipeline to interleave them with loads, achieving effective ILP of 3.5 across 4 VPUs. The final \texttt{argmax} over the $H \times W$ score map is handled entirely by the Ibex core.

\subsection{RVV 1.0 Implementation and Performance Scaling}
\label{subsec:rvv_impl}
We model VFT execution on HyperV under MAP encoding with 16-bit integer storage. With $\mathrm{VLEN}=4096$, $\mathrm{SEW}=16$, and $\mathrm{LMUL}=1$, each register holds 256 elements, allowing (D=64) through (D=256) to be processed in one vector instruction. Execution time therefore scales linearly with (D) and quadratically with the spatial field width.

The four-VPU backend exposes substantial parallelism across independent cells. Basis shifts and local writes approach an ILP of 4.0, while dependency chains in basis mixing and reduction latency lower the corresponding ILP to approximately 3.2 and 3.5. We convert vector beats to the cycle estimates in Table~\ref{tab:kernel_perf} using an assumed 67\% average utilization that accounts for bank contention, pipeline startup, and queue draining.

For (H=W=15), (D=256), and (K=1), one step requires 112,300 beats and an estimated 41,902 cycles at 200 MHz. Basis mixing accounts for 54.5\% of the beats, followed by shifting at 30.0\% and querying at 15.2\%. Across the evaluated configurations, latency remains below 1 ms, motivating specialized support for the data-movement-heavy shift kernel.

\begin{table}[t]
\caption{VFT kernel performance on HyperV at 200\,MHz. All configurations use MAP encoding with 16-bit signed integer storage. Vector instructions and beats are per single VFT step (transport $+$ write $+$ query).}
\label{tab:kernel_perf}
\vspace{-2mm}
\centering
\renewcommand{\arraystretch}{1.15}
\setlength{\tabcolsep}{3.5pt}
\resizebox{\columnwidth}{!}{%
\begin{tabular}{c c c r r r r r r}
\toprule
\textbf{Grid} & \textbf{$D$} & \textbf{$K$} & \textbf{Vec Insns} & \textbf{Beats} & \textbf{Cycles} & \textbf{Latency} & \textbf{Memory} & \textbf{ILP} \\
$H{\times}W$ & & & & & (est.) & ($\mu$s) & (kB) & (avg) \\
\midrule
$8{\times}8$    & 64  & 1 & 1,609  & 7,908   & 2,950   & 14.8  & 8.0   & 3.48 \\
$8{\times}8$    & 128 & 1 & 3,218  & 15,816  & 5,901   & 29.5  & 16.0  & 3.48 \\
$8{\times}8$    & 256 & 1 & 6,436  & 31,632  & 11,802  & 59.0  & 32.0  & 3.48 \\
\midrule
$9{\times}9$    & 64  & 1 & 2,041  & 10,027  & 3,740   & 18.7  & 10.1  & 3.48 \\
$9{\times}9$    & 128 & 1 & 4,082  & 20,054  & 7,482   & 37.4  & 20.2  & 3.48 \\
$9{\times}9$    & 256 & 1 & 8,164  & 40,108  & 14,965  & 74.8  & 40.5  & 3.48 \\
\midrule
$15{\times}15$  & 64  & 1 & 5,725  & 28,075  & 10,474  & 52.4  & 28.1  & 3.49 \\
$15{\times}15$  & 128 & 1 & 11,450 & 56,150  & 20,950  & 104.8 & 56.2  & 3.49 \\
$15{\times}15$  & 256 & 1 & 22,900 & 112,300 & 41,902  & 209.5 & 112.5 & 3.49 \\
$15{\times}15$  & 256 & 4 & 53,500 & 263,500 & 98,320  & 491.6 & 450.0 & 3.25 \\
\midrule
$32{\times}32$  & 64  & 1 & 26,329 & 128,868 & 48,085  & 240.4 & 128.0 & 3.49 \\
$32{\times}32$  & 256 & 1 & 105,316& 515,472 & 192,340 & 961.7 & 512.0 & 3.49 \\
\bottomrule
\end{tabular}%
}
\vspace{-4mm}
\end{table}

\subsection{Field Shift DMA Engine}
\label{subsec:dma}

\begin{figure*}
    \centering
    \includegraphics[width=1\textwidth]{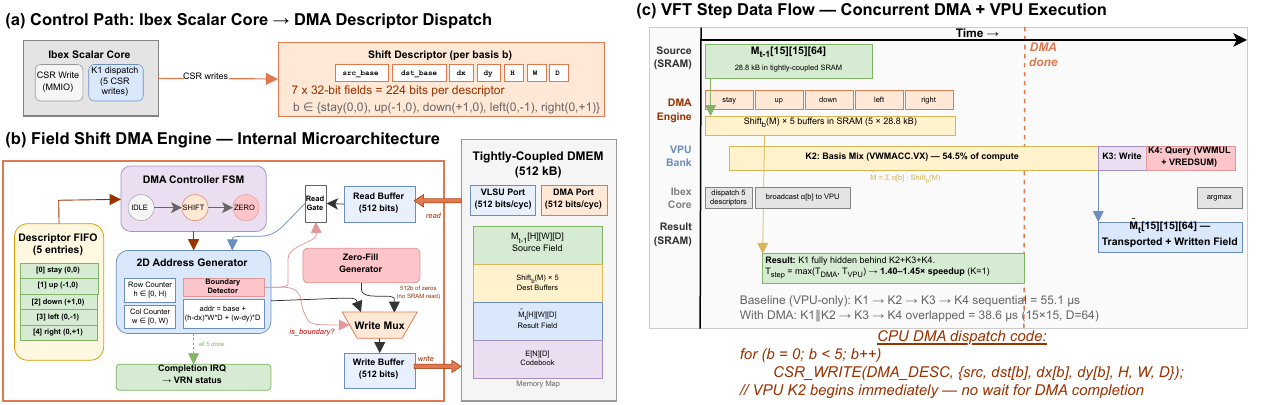}
    \caption{Micro-architecture and execution flow of the Field Shift DMA Engine. (a) The scalar core dispatches shift descriptors to initiate memory transfers. (b) The DMA manages 2D address generation and boundary zero-filling, offloading pure data movement from the VPUs. (c) Temporal overlap enables spatial shifts to execute concurrently with VPU arithmetic, hiding data movement latency.}
    \label{fig:dma_arch}
    \vspace{-4mm}

\end{figure*}

Kernel profiling in Table~\ref{tab:kernel_perf} reveals that the basis shift kernel (K1) consistently accounts for 29--30\% of total beats across all single-channel configurations. This kernel performs no arithmetic---it is pure structured data movement, copying $(H{-}|dx|)(W{-}|dy|)$ contiguous $D$-element cell vectors to displaced addresses and zeroing vacated boundary cells. Executing these shifts on general-purpose VPUs wastes compute resources and instruction queue capacity, displacing the arithmetic-heavy basis mix and query kernels.

We address this by introducing a \textbf{Field Shift DMA Engine}, a lightweight fixed-function unit that executes all five basis shifts concurrently with VPU computation, removing K1 from the critical path entirely. The DMA engine is positioned as an independent hardware unit configured via software based control status register (CSR) accesses, connecting to the tightly coupled SRAM through a dedicated 512-bit read/write port.

\textbf{Microarchitecture.}
The DMA engine comprises three components. A \textit{descriptor FIFO} (5 entries, one per basis direction) receives shift descriptors from the Ibex core via memory-mapped CSR writes. Each descriptor encodes $(\texttt{src\_base}, \texttt{dst\_base}, dx, dy, H, W, D)$ in 7 scalar registers. A \textit{2D address generator} consumes descriptors and emits source-destination address pairs, iterating over rows and columns with $(dx, dy)$ displacement. The generator skips boundary cells mapping to vacated regions, directing those addresses to a \textit{zero-fill unit} that writes pre-initialized zero vectors without issuing a read.

Each transfer copies one $D$-element cell vector per transaction. At 512\,bits/cycle on the dedicated SRAM port, a cell transfer requires $D \cdot 16 / 512$ cycles (2 cycles for $D{=}64$, 8 for $D{=}256$). Since all five basis shifts are independent, the DMA processes them sequentially from the FIFO; total DMA latency for one transport step is
\begin{equation}
\label{eq:dma_cycles}
T_\text{DMA} = |\mathcal{B}| \cdot \left[(H{-}1)(W{-}1) + \max(H,W)\right] \cdot \frac{D \cdot 16}{512},
\end{equation}
where $|\mathcal{B}|{=}5$ and the $\max(H,W)$ term accounts for boundary zero-fill operations.

\textbf{Concurrent execution model.}
The scalar core dispatches five shift descriptors at the start of each transport step and immediately begins issuing basis mix (K2) instructions to the VPU bank. The DMA produces shifted field buffers in SRAM while the VPUs execute K2 on already-available portions. Upon completion, the DMA raises an interrupt and VPUs proceed to local write (K3) and query (K4) without synchronization stalls. The total latency is therefore
\begin{equation}
\label{eq:dma_overlap}
T_\text{step} = \max\!\left(T_\text{DMA},\; T_\text{VPU}^{\text{K2+K3+K4}}\right),
\end{equation}
where $T_\text{VPU}^{\text{K2+K3+K4}}$ is the VPU execution time for the remaining kernels. Since $T_\text{DMA}$ is bandwidth-dominated and $T_\text{VPU}$ is compute-dominated, the DMA is fully hidden behind VPU execution in all practical configurations.

\textbf{Performance evaluation.}
Table~\ref{tab:dma_perf} reports VFT step latency with and without the DMA engine. For non-temporal configurations ($K{=}1$), the DMA provides a consistent \textbf{1.40--1.45$\times$ speedup} by eliminating K1 from the VPU critical path. The egomotion configuration ($15{\times}15$, $D{=}64$) improves from 52.4\,$\mu$s to 36.7\,$\mu$s per step. For the full temporal configuration ($K{=}4$), additional VPU work from age-field and temporal aggregation kernels extends the VPU critical path, reducing the relative benefit to 1.15$\times$; nevertheless, the DMA still saves over 63\,$\mu$s per step. At the largest configuration ($32{\times}32$, $D{=}256$), latency drops from 962\,$\mu$s to 665\,$\mu$s, keeping VFT well within real-time budgets.

\textbf{Area overhead.}
The DMA engine requires a 5-entry descriptor FIFO ($5 \times 7 \times 32 = 1{,}120$\,bits), a 2D address counter (two 16-bit counters plus comparators), zero-fill gating logic, and a dedicated SRAM port arbiter. We estimate total area under 0.02\,mm$^2$ in 28\,nm, less than 1\% of HyperV's 2.7\,mm$^2$ die area.

\begin{table}[t]
\caption{VFT step latency with and without the Field Shift DMA engine. The DMA executes all five basis shifts concurrently with VPU compute, eliminating K1 from the critical path.}
\label{tab:dma_perf}
\vspace{-2mm}
\centering
\renewcommand{\arraystretch}{1.15}
\setlength{\tabcolsep}{3pt}
\resizebox{\columnwidth}{!}{%
\begin{tabular}{c c c r r r r c}
\toprule
\textbf{Grid} & $D$ & $K$ & \textbf{Baseline} & \textbf{w/ DMA} & \textbf{DMA} & \textbf{VPU} & \textbf{Speedup} \\
$H{\times}W$ & & & ($\mu$s) & ($\mu$s) & (cyc) & (cyc) & \\
\midrule
$8{\times}8$    & 64  & 1 & 14.8  & 10.5  & 570    & 2,107   & 1.41$\times$ \\
$8{\times}8$    & 128 & 1 & 29.5  & 21.1  & 1,140  & 4,214   & 1.40$\times$ \\
$8{\times}8$    & 256 & 1 & 59.0  & 42.1  & 2,280  & 8,429   & 1.40$\times$ \\
\midrule
$9{\times}9$    & 64  & 1 & 18.7  & 13.3  & 730    & 2,659   & 1.41$\times$ \\
$9{\times}9$    & 128 & 1 & 37.4  & 26.6  & 1,460  & 5,318   & 1.41$\times$ \\
$9{\times}9$    & 256 & 1 & 74.8  & 53.2  & 2,920  & 10,637  & 1.41$\times$ \\
\midrule
$15{\times}15$  & 64  & 1 & 52.4  & 36.7  & 2,110  & 7,333   & 1.43$\times$ \\
$15{\times}15$  & 128 & 1 & 104.8 & 73.3  & 4,220  & 14,667  & 1.43$\times$ \\
$15{\times}15$  & 256 & 1 & 209.5 & 146.7 & 8,440  & 29,335  & 1.43$\times$ \\
$15{\times}15$  & 256 & 4 & 491.6 & 428.8 & 8,440  & 85,753  & 1.15$\times$ \\
\midrule
$32{\times}32$  & 64  & 1 & 240.4 & 166.4 & 9,930  & 33,271  & 1.44$\times$ \\
$32{\times}32$  & 256 & 1 & 961.7 & 665.4 & 39,720 & 133,086 & 1.45$\times$ \\
\bottomrule
\end{tabular}%
}
\vspace{-4mm}
\end{table}

\section{Results}
\subsection{Experimental Setup}
We compare VFT against flat VSA memories using identical token encodings. To isolate the effect of memory organization, all results are reported for both MAP and FHRR instantiations. Unless otherwise stated, the default hypervector dimension is $D=256$, results are averaged over five random seeds, and we report both top-1 recall and mean $\ell_1$ position error.

\begin{figure*}
    \centering
    \includegraphics[width=\linewidth]{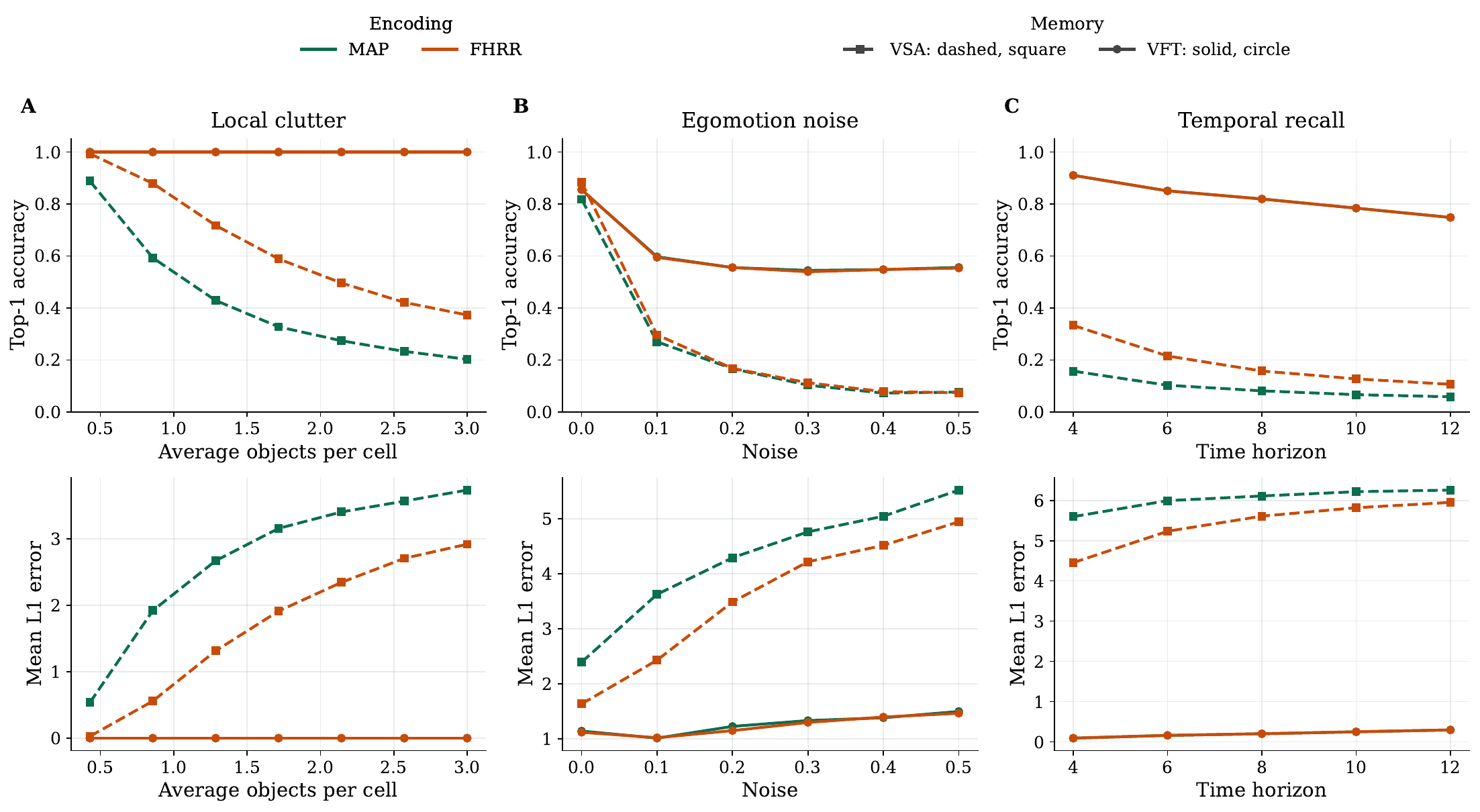}
    \caption{Spatio-temporal benchmarks comparing VSA vs VFT memory. A) Local clutter: VFT significantly outperforms VSA as number of objects in the visual scene increases. B) Egomotion noise: VFT accuracy degrades with noise but stabilizes whereas VSA continues to decrease to below 10\% accuracy. C) Temporal Recall: VFT accuracy remains consistently high across different time horizons regardless of encoding method whereas VSA struggles at even time horizon = 4.}
    \label{fig:main}
    \vspace{-4mm}
\end{figure*}

\subsection{Benchmarks}
\textbf{Local clutter.} Objects are written into a $8\times8$ field with replacement so that multiple objects collide in the same cell. The number of objects is swept from 21 to 147, corresponding to 0.5-3.0 objects per cell on average. This benchmark tests whether spatial indexing mitigates the unbinding interference that limits flat VSA memories under heavy local superposition. 

\textbf{Egomotion noise.} An agent moves in a $32\times32$ world while observing 18 objects. The maintained VFT field is $9\times9$, rollouts last 24 steps and noise is swept from 0.0 to 0.5. The task is to recover final relative object positions after repeated motion and observation updates. This benchmark evaluates whether VFT is robust under noisy conditions.

\textbf{Temporal horizon.} We use a $15\times15$ grid with 32 objects, four age channels, and blending factor 0.75 while sweeping the recall horizon from 4 to 12 steps. Queries ask for object positions at earlier timesteps. This benchmark measures whether temporal history can be retained through channelized memory organization instead of explicit time-token binding. 

\textbf{Rollout length.} Using a $32\times32$ world, a $9\times9$ field, and 24 objects, rollout length is swept from 4 to 48 steps. We compare deterministic transport against probabilistic transport based on the action confusion model. This test measures whether uncertainty-aware transport reduces compounding drift over long horizons.

\begin{figure}
    \centering
    \includegraphics[width=\linewidth]{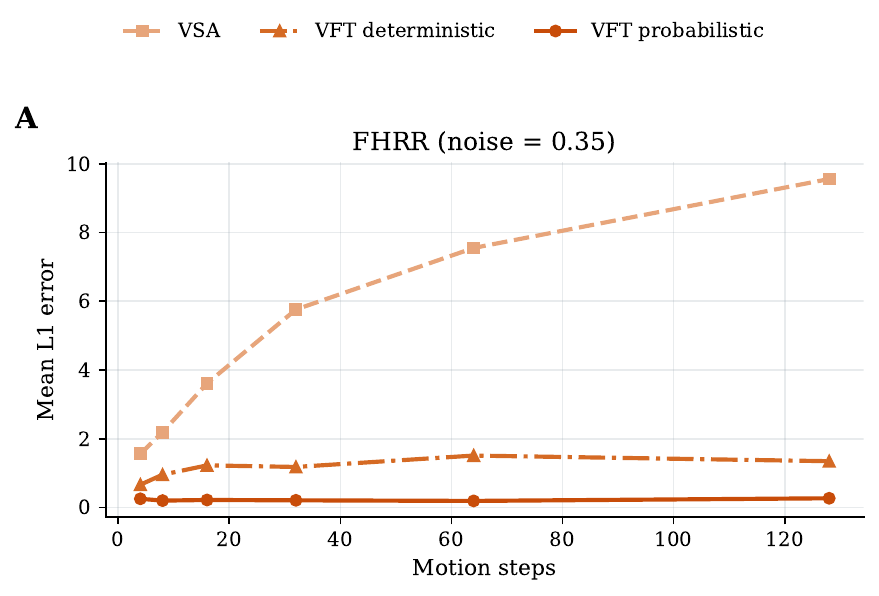}
    \caption{Rollout Experiment testing the performance of VSA, VFT-Deterministic, and VFT-Probabilistic between 4 to 128 steps. VFT-probabilistic has the lowest Mean L1 score whereas VSA's error continues to rise as the number of steps increases.}
    \label{fig:rollout_l1}
    \vspace{-5mm}
\end{figure}

\begin{figure*}
    \centering
    \includegraphics[width=\linewidth]{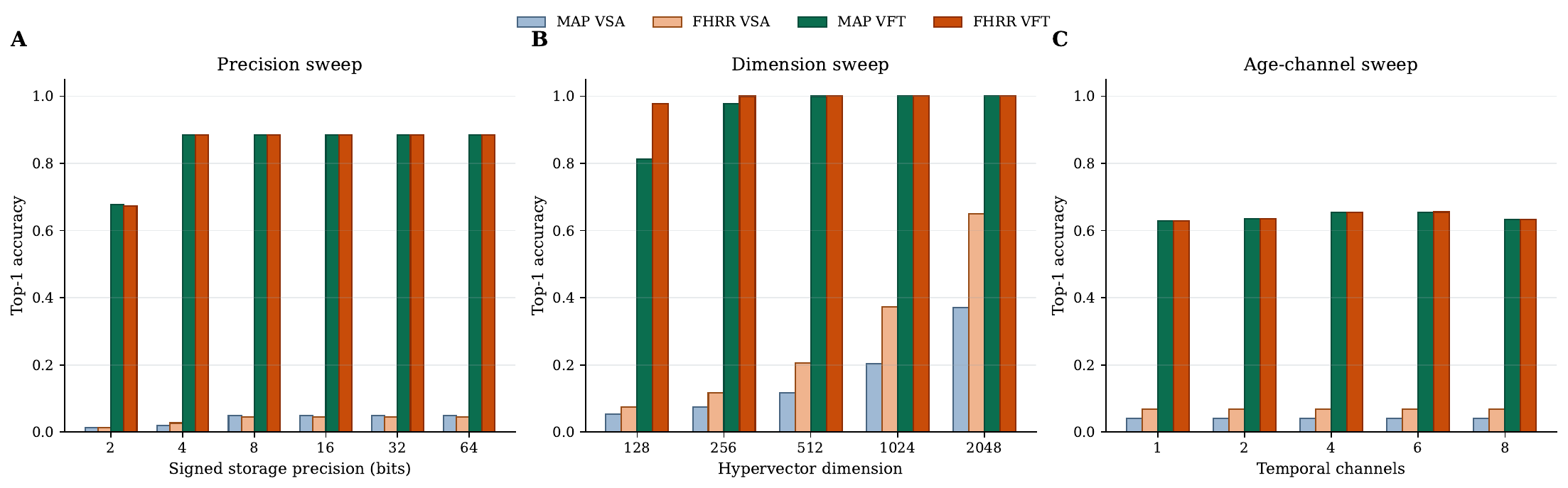}
    \caption{Ablation studies comparing Flat VSA memory vs VFT utilizing both MAP and FHRR encodings. A) Bit precision sweep between 2-bit to 64-bit B) Dimensionality sweep between $D = 128 - 2048$. C) Number of $K$ Channels sweep between 1 to 8 channels.}
    \label{fig:ablation}
    \vspace{-5mm}
\end{figure*}

\subsection{Main Benchmark Results}
Figure~\ref{fig:main} summarize the main results. Across all three primary benchmarks, VFT substantially outperforms flat VSA for both MAP and FHRR, indicating that the gains come from memory organization rather than the choice of encoding.

At the hardest clutter setting, VFT achieves 100.0\% top-1 accuracy for both MAP and FHRR, while flat VSA drops to 20.3\% and 37.3\%, yielding gains of 79.7 and 62.7 points. This shows that the dominant failure mode in cluttered scenes is not the encoding itself, but the use of a single flat superposition memory. Spatial indexing localizes interference and preserves exact recall.

In Figure~\ref{fig:main}B, under 50\% noise, VFT reaches 55.6\% accuracy for MAP and 55.4\% for FHRR, while flat VSA collapses to 7.7\% and 7.4\%. The roughly 48-point improvement shows that transporting a structured egocentric field is far more robust to motion uncertainty than repeatedly rebinding observations into a global flat memory.

Additionally, in Figure~\ref{fig:main}C, we see that at horizon 12, VFT achieves 74.8\% accuracy for both encodings, whereas flat temporal VSA reaches only 5.8\% for MAP and 10.6\% for FHRR. The gains of 69.0 and 64.2 points show that explicit age channels preserve useful temporal history far better than binding time directly into a flat superposition.

For all experiments, we notice a general trend that the FHRR encoding helps flat VSA memory performance, yet still flat VSA memory is not comparable to VFT memory architectures in terms of accuracy or $\ell_1$ error. The three benchmarks point to the same conclusion. VFT reduces spatial interference, maintains alignment under noisy motion, and preserves temporal history in dynamic scenes. The consistency across MAP and FHRR indicates that the benefit is architectural.

\subsection{Long-Horizon Rollout Behavior}
Figure~\ref{fig:rollout_l1} evaluates transport over long noisy rollouts. The probablistic transport significantly outperforms the VSA model and outperforms the deterministic transport as well. This shows that once the model is uncertain, repeatedly applying a single inverse shift accumulates substantial drift, whereas distributing mass across plausible inverse motions better preserves recoverable state. This result highlights that transport is a meaningful and improvable component of the VFT memory update rule, and is essentialy for future VSA works revolving around spatial-temporal tasks.

\subsection{Ablation Results}
We additionally report three ablations consisting of precision (2-64 bits), hypervector dimension ($D=128-2048$), and channel count (1-8) to see the effects that quantization, dimensionality, and temporal channels have on the VFT model for edge device applicability. Both MAP and FHRR VFT reach 95\% of their 64-bit egomotion accuracy with only 4-bit storage, indicating that the main advantage comes from memory structure rather than numeric precision. The largest VFT-VSA gap in local clutter occurs at $D=256$ for MAP and $D=128$ for FHRR. Thus, the benefit of VFT appears well before very high dimension, and increasing dimension alone does not rescue flat superposition memories under heavy clutter whereas FHRR-VSA at $D=2048$ still struggles to achieve the same accuracy as VFT at $D=128$. A single channel reaches 95\% of the best temporal performance for both encodings under the tested setting, suggesting diminishing returns from additional channels in this regime. For the highest accuracy, we have chosen to use 4 temporal channels.

\section{Conclusion}
This paper introduces Vector Field Transforms, a structured symbolic spatio-temporal memory that addresses a core limitation of conventional flat VSA memory. By replacing a single global superposition with a spatially indexed field, updating that field with an explicit transport operator, and supporting temporal recall through age channels, VFT enables symbolic memory to evolve coherently under motion. Across MAP and FHRR, VFT consistently improves spatial, egocentric, and temporal recall over flat VSA baselines, showing that memory organization is essential for symbolic world-state reasoning. We further showed that VFT generates a medium-scale vector workload that maps naturally to a novel programmable RVV processor such as HyperV. Taken together, these results position VFT as a practical bridge between hardware-efficient symbolic computation and structured spatio-temporal memory for edge intelligence.

\section{Acknowledgment}
This material is based upon work supported by the Air Force Office of Scientific Research under award number FA9550-23-F-0014 in the amount of \$129600. This work was also supported in part by the DARPA Young Faculty Award, the National Science Foundation (NSF) under Grants \#2127780, \#2319198, \#2321840, \#2312517, and \#2235472, the Semiconductor Research Corporation (SRC), the Office of Naval Research through the Young Investigator Program Award, and Grants \#N00014-21-1-2225 and \#N00014-24-1-2547, Army Research Office Grant \#W911NF2410360. Additionally, support was provided by the Air Force Office of Scientific Research under Award \#FA9550-22-1-0253.

\newpage
{
\bibliographystyle{IEEEtran}
\bibliography{mybibliography}
}

\end{document}